\documentclass[a4paper,fleqn]{cas-dc}

\def\tsc#1{\csdef{#1}{\textsc{\lowercase{#1}}\xspace}}
\tsc{TK}
\tsc{AD}
\tsc{KDL}
\begin{document}
\let\WriteBookmarks\relax
\def\floatpagepagefraction{1}
\def\textpagefraction{.001}
\shorttitle{Subdiffusion equation with Cattaneo effect}    
\shortauthors{T.~Koszto{\l}owicz, A.~Dutkiewicz, K.D.~Lewandowska}  

\title [mode = title]{Subdiffusion equation with Cattaneo effect}  



\author[1,2]{Tadeusz Koszto{\l}owicz}[orcid=0000-0001-5710-2970]

\cormark[1]


\ead{tadeusz.kosztolowicz@ujk.edu.pl}



\affiliation[1]{organization={Institute of Physics, Jan Kochanowski University},
             addressline={Uniwersytecka 7},
              postcode={25-406}, 
            city={Kielce},
%          citysep={}, % Uncomment if no comma needed between city and postcode
            country={Poland}}

\affiliation[2]{organization={Department of Radiological Informatics and Statistics, Medical University of Gda\'nsk},
  addressline={Tuwima 15},
             postcode={80-210},  
            city={Gda\'nsk},
%          citysep={}, % Uncomment if no comma needed between city and postcode
            country={Poland}}

\author[3]{Aldona Dutkiewicz}[orcid=0000-0002-5171-4009]

% Footnote of the second author

% Email id of the second author
\ead{szukala@amu.edu.pl}



\affiliation[3]{organization={Faculty of Mathematics and Computer Science, Adam Mickiewicz University},
  addressline={Uniwersytetu Pozna\'nskiego~4},
              postcode={61-614}, 
            city={Pozna\'n},
%          citysep={}, % Uncomment if no comma needed between city and postcode
            country={Poland}}

\author[4]{Katarzyna D. Lewandowska}[orcid=0000-0001-8883-1156]

% Footnote of the third author
%\fnmark[3]

% Email id of the third author
\ead{kale@gumed.edu.pl}

% URL of the third author


\affiliation[4]{organization={Department of Physics and Biophysics, Medical University of Gda\'nsk},
  addressline={D\k{e}binki 1},
              postcode={80-211}, 
            city={Gda\'nsk},
%          citysep={}, % Uncomment if no comma needed between city and postcode
            country={Poland}}

\cortext[cor1]{Corresponding author}


\begin{abstract}
The ordinary subdiffusion equation, with a fractional time derivative of at most first order, describes a process in which the propagation velocity of diffusing molecules is unlimited. To avoid this non-physical property different forms of the Cattaneo subdiffusion equation have been proposed. 
We define the Cattaneo effect as a delay of the ordinary subdiffusion flux activation by a random time. 
By incorporating this effect into the flux equation we get a Cattaneo--type subdiffusion equation (CTSE).
We consider a subdiffusion process in which the Cattaneo effect is generated by the time-delay probability distribution controlled by the Mittag-Leffler function. Then, CTSE differs from the ordinary subdiffusion equation by a term with a fractional time derivative, whose order can be independent of the subdiffusion exponent. The influence of the Cattaneo effect on the solutions to the CTSE is discussed. 
We show that the process described by CTSE is subdiffusion in the entire time domain even though the temporal evolution of the mean square displacement of diffusing particle in the short-time limit is typical for superdiffusion. The delay in the flux activation in the subdiffusion equation should also cause a flux delay in a boundary condition. As an example, we study subdiffusion with the Cattaneo effect in a system with a partially absorbing wall at which the Robin boundary condition is assumed. We also propose a method for experimentally identifying the Cattaneo effect in a subdiffusive system. 
\end{abstract}


\begin{highlights}
\item Definition of the Cattaneo effect.
\item Derivation of the Cattaneo-type subdiffusion equation.
\item Boundary condition with the Cattaneo effect.
\item Green's function for a system with partially absorbing wall.
\item Time evolution of the mean square displacement of a diffusing particle.
\end{highlights}

%\nocite{*}

% Keywords

\maketitle

\section{Introduction\label{secI}}

A type of diffusion is often defined by the temporal evolution of the mean squared displacement (MSD) $\sigma^2(t)$ of a randomly wandering particle. In normal diffusion, $\sigma^2$ grows linearly with time, $\sigma^2(t)\sim t$, deviations from this formula correspond to anomalous diffusion. Superdiffusion is defined as a process which is significantly faster than normal diffusion, then anomalously long particle jumps can be performed with high probability. It occurs, among others, in turbulent media, in random velocity fields \cite{redner1989,zumofen1990,bouchaud1990,compte1998}, movement of soil amebas on glass surfaces \cite{levandowsky1997}, mussels movement \cite{dejager2011}, and cell migration in biological processes \cite{dieterich2008}. 
Subdiffusion occurs in media in which particle random walk is very hindered. The examples are transport of molecules in viscoelastic chromatin network \cite{lee2021}, in porous media \cite{bijeljic2011}, and in living cells \cite{barkai2012}, transport of sugars in agarose gel \cite{kdm}, and antibiotics in bacterial biofilm \cite{km}. In this case, the waiting time for a particle next jump is anomalously long. The most commonly considered function is $\sigma^2(t)\sim t^\alpha$, then $\alpha>1$ is for superdiffusion and $0<\alpha<1$ for subdiffusion. 
We consider subdiffusion in a one-dimensional homogeneous system with constant parameters describing the process.

Frequently, ordinary subdiffusion is described by a fractional differential equation with the Riemann-Liouville fractional time derivative of the order $1-\alpha$ or with the Caputo time derivative of the order $\alpha$ \cite{wyss1986,hilferanton,compte,mks,mk,barkai2000,skb,klages2008,ks}. 
For the parabolic normal diffusion equation and subdiffusion equation with a time derivative of at most first order the probability density of finding the molecule at position $x$ at time $t$ (the Green's function, GF) $P(x,t|x_0)$ is greater than zero for any $x$ and $t>0$, $x_0$ is the initial particle position. It means that the propagation velocity of a particle is unlimited. To avoid this non--physical property, the Cattaneo normal diffusion equation has been proposed; the equation was originally used to describe heat propagation \cite{cattaneo}. The equation involves a second-order time derivative. 

While the Cattaneo hyperbolic normal diffusion equation is well defined, the Cattaneo subdiffusion equation takes various non-equivalent forms, see Refs. \cite{compte1997,fernandez,qi,liu,roscani,alegria,luchko,awad2021,awad2019,hamada,metzler1999,metzler1998,gorska2020,gorska2021,koszt2014,masoliver,masoliver2016,povstenko2011}. The Cattaneo equation with the Caputo and/or the Riemann--Liouville fractional derivatives have been used to describe subdiffusion of neutrons inside the core of a nuclear reactor \cite{vya}, heat transport in porous media \cite{nikan}, and in a system with glass spheres in a tank filled with air \cite{moza}. Cattaneo--type equations with fractional derivatives other than those mentioned above have been also considered. The examples are the equations with Caputo--Fabrizio fractional time derivative \cite{liu2017}, with tempered Caputo derivative \cite{beghin}, and with Hilfer fractional derivative with respect to another function \cite{vieira}. 

If the diffusion properties of the system do not change over time, the type of diffusion, as well as parameters describing the process, should not change either (assuming that the parameters do not depend on the concentration of diffusing substance). However, mathematical properties of solutions to the Cattaneo-type subdiffusion equation are in some contradiction with the above-mentioned physical condition.
When diffusion in a subdiffusive medium is described by the Cattaneo--type equation, for very short time there is $\sigma^2(t)\sim t^\nu$ with $\nu>1$. For example, there are $\nu=2\alpha$, $\nu=1+\alpha$, and $\nu=2$ for different models considered in Ref. \cite{compte1997}. The relation $\sigma^2(t)\sim t^\alpha$ with $\alpha\in(0,1)$, typical for subdiffusion, appears for longer times. It seems to be unphysical that superdiffusion could occur in a subdiffusive medium whose diffusion properties do not change over time. 

Solutions to the Cattaneo--type subdiffusion equation (CTSE) and solutions to the subdiffusion equation without the Cattaneo effect are usually not much different from each other. Then, the latter equation, which appears to be easier to solve, is used to describe subdiffusion. However, there are processes that both equations provide qualitatively different results even when the Cattaneo effect is small. Example of this is the diffusive and subdiffusive impedance \cite{barbero,kostrobij,kl2009,lk2008}. 
The question arises whether CTSE provides significantly different results than the ordinary subdiffusion equation. 

We define the Cattaneo effect as a delay of the activation of ordinary subdiffusion flux by a random time. Combining the constitutive equation, which describes the relation between the probability density flux and the gradient of Green's function, with the continuity equation the Cattaneo--type subdiffusion equation is obtained. We consider the CTSE that differs from the ordinary subdiffusion equation by an additional term with the fractional time derivative of the order larger than $\alpha$. 
We show that the process described by the subdiffusion equation with the Cattaneo effect can be interpreted as subdiffusion even for very short times. Next, we consider the influence of the Cattaneo effect not only on the equation but also on the boundary conditions. As an example, we consider subdiffusion in a system with a partially absorbing wall. 

The article is organized as follows. In Sec. \ref{secII}, the definition and interpretation of the Cattaneo effect and the general form of the CTSE are presented. Next, the subdiffusion with Cattaneo effect generated by the Mittag-Leffler function is considered. The CTSE, along with the Green's functions for the unbounded system are obtained. The functions are derived separately for short and long times. A comparison of the Green's function and the time evolution of the MSD for subdiffusion processes with and without the Cattaneo effect are presented in Sec. \ref{secIII}. In Sec. \ref{secIV}, we consider subdiffusion with the Cattaneo effect in a system with a partially absorbing wall. Concluding remarks are presented in Sec. \ref{secV}.

\section{Subdiffusion equation with Cattaneo effect\label{secII}}

In this section, we define the Cattaneo effect, derive the subdiffusion equation taking this effect into account, and find the Green's function for the unbounded system.

\subsection{General form of Cattaneo--type subdiffusion equation\label{secIIa}}

In the following, we denote functions describing a process in which the Cattaneo effect does not occur by $\tau=0$, the interpretation of the parameter $\tau$ will be presented later in this section.
The ordinary subdiffusion equation, derived within the continuous time random walk model \cite{mk,ks}, reads
\begin{equation}\label{eqIIa1}
\frac{\partial P_{\tau=0}(x,t|x_0)}{\partial t}=D\frac{^{RL}\partial^{1-\alpha}}{\partial t^{1-\alpha}}\frac{\partial^2 P_{\tau=0}(x,t|x_0)}{\partial x^2},
\end{equation}
where $\alpha\in(0,1)$ is the subdiffusion parameter (exponent), and $D$ is a subdiffusion coefficient given in units of ${\rm m^2/sec^\alpha}$. The Riemann--Liouville fractional derivative, involved in Eq. (\ref{eqIIa1}), is defined as
\begin{equation}\label{eqIIa2}
\frac{^{RL}d^\nu f(t)}{dt^\nu}=\frac{1}{\Gamma(n-\nu)}\frac{d^n}{dt^n}\int_0^t (t-u)^{n-\nu-1}f(u)du,
\end{equation}
where $\nu>0$ and $n$ is a natural number, $n=\lfloor\nu\rfloor+1$ when $\nu\notin\mathcal{N}$, otherwise $n=\nu\in\mathcal{N}$.

A diffusion equation can be obtained phenomenologically by a combination of the continuity equation
\begin{equation}\label{eqIIa3}
\frac{\partial P(x,t|x_0)}{\partial t}=-\frac{\partial J[P(x,t|x_0)]}{\partial x},
\end{equation}
and a constitutive equation defining the flux, $J[P]$ is the flux operator acting on the function $P$.  
For ordinary subdiffusion the constitutive flux equation is defined as
\begin{equation}\label{eqIIa4}
J_{\tau=0}[P_{\tau=0}(x,t|x_0)]=-D\frac{^{RL}\partial^{1-\alpha}}{\partial t^{1-\alpha}}\frac{\partial P_{\tau=0}(x,t|x_0)}{\partial x}.
\end{equation}
We define the Cattaneo effect as the ordinary subdiffusive flux activation delay by a random time, 
\begin{equation}\label{eqIIa5} 
J[P(x,t|x_0)]=\int_0^t R(t';\tau)J_{\tau=0}[P(x,t-t'|x_0)]dt',
\end{equation}
where $R$ is a probability density of the delay time. The Cattaneo--type subdiffusion equation is obtained by combination of Eqs. (\ref{eqIIa3}) and (\ref{eqIIa5}). 

When analyzing diffusion equations it is convenient to use the Laplace transform $\mathcal{L}[f(t)](s)=\int_0^\infty {\rm e}^{-st}f(t)dt\equiv \hat{f}(s)$. In the calculations, we use the relations
$\mathcal{L}[f^{(n)}(t)](s)=s^n\hat{f}(s)-\sum_{j=0}^{n-1}s^jf^{(n-1-j)}(0)$, where $f^{(n)}(t)=d^n f(t)/dt^n$, $\mathcal{L}[(f\ast h)(t)](s)=\hat{f}(s)\hat{h}(s)$, where $(f\ast h)(t)=\int_0^t f(t')g(t-t')dt'$,
and Ref. \cite{ks}, p.82,
\begin{equation}\label{eqIIa6}
\mathcal{L}\Bigg[\frac{^{RL}d^\alpha f(t)}{dt^\alpha}\Bigg](s)=s^\alpha \mathcal{L}\left[f(t)\right](s),\;\alpha\in(0,1).
\end{equation} 
The LT of Eqs. (\ref{eqIIa3})--(\ref{eqIIa5}) are, respectively,
\begin{equation}\label{eqIIa7}
s\hat{P}(x,s|x_0)-P(x,0|x_0)=-\frac{\partial \hat{J}[\hat{P}(x,s|x_0)]}{\partial x},
\end{equation}
\begin{equation}\label{eqIIa8}
\hat{J}_{\tau=0}[\hat{P}_{\tau=0}(x,s|x_0)]=-Ds^{1-\alpha}\frac{\partial\hat{P}_{\tau=0}(x,s|x_0)}{\partial x},
\end{equation}
and
\begin{equation}\label{eqIIa9}
\hat{J}[\hat{P}(x,s|x_0)]=\hat{R}(s;\tau)\hat{J}_{\tau=0}[\hat{P}(x,s|x_0)].
\end{equation}

Combining Eqs. (\ref{eqIIa7})--(\ref{eqIIa9}) we obtain
\begin{equation}\label{eqIIa10}
s\hat{P}(x,s|x_0)-P(x,0|x_0)=D\hat{R}(s;\tau)s^{1-\alpha}\frac{\partial^2 \hat{P}(x,s|x_0)}{\partial x^2}.
\end{equation}
In the time domain Eq. (\ref{eqIIa10}) reads
\begin{equation}\label{eqIIa11}
\mathcal{R}_{\tau}[P(x,t|x_0)]
=D\frac{^{RL}\partial^{1-\alpha}}{\partial t^{1-\alpha}}\frac{\partial^2 P(x,t|x_0)}{\partial x^2},
\end{equation}
where
\begin{equation}\label{eqIIa12}
\mathcal{R}_{\tau}[f(t)]=\int_0^t \mathcal{L}^{-1}\left[\frac{1}{\hat{R}(s;\tau)}\right](t')f^{(1)}(t-t')dt'.
\end{equation}

The Cattaneo effect is controlled by the probability distribution $R$ which depends on the parameter $\tau$. We assume that the Cattaneo effect is turned off when $\tau=0$.
The function $R$ satisfies the following conditions: 
\begin{enumerate}
	\item $R(t;\tau)$ is normalized, $\int_0^\infty R(t;\tau)dt=1\Leftrightarrow\hat{R}(0;\tau)=1$,
	\item $R(t;\tau)$ is non-negative in the entire time domain. According to the Bernstein theorem, this property is satisfied when $\hat{R}(s;\tau)$ is a completely monotonic function ($\mathcal{CMF}$), i.e. it satisfies the condition \\
	$(-1)^n \partial^n\hat{R}(s;\tau)/\partial s^n\geq 0$ for $n=0,1,2,\ldots$ \cite{gorska2020,schilling}, 
	\item the Cattaneo effect disappears when $\tau=0$, then $R(t;0)=\delta(t)\Leftrightarrow\hat{R}(s;0)=1$, where $\delta$ is the delta--Dirac function.  
\end{enumerate}

\subsection{Cattaneo effect generated by the Mittag--Leffler function\label{secIIb}}

In the following we consider the function $R$ for which the Laplace transform is  
\begin{equation}\label{eqIIb1}
\hat{R}(s;\tau)=\frac{1}{1+\tau s^\kappa},
\end{equation} 
where $\kappa\in(0,1)$, which ensures that $\hat{R}\in\mathcal{CMF}$ \cite{gorska2020,schilling}, $\tau$ is given in the units of $\rm{sec}^\kappa$.
The determination of the inverse Laplace transform is performed separately for large and small parameter $s$. Assuming $\tau s^\kappa>1$ we get the inverse transform for short time, when $\tau s^\kappa<1$ we find $R(t;\tau)$ for long time. Using the method described in the Appendix we obtain:
\begin{itemize}
	\item for short times
	\begin{equation}\label{eqIIb2}
	R(t;\tau)=\frac{1}{\tau t^{1-\kappa}}E_{\kappa,\kappa}\left(-\frac{t^\kappa}{\tau}\right),
	\end{equation}
	where 
	\begin{equation}\label{eqIIb3}
	E_{\rho,\beta}(u)=\sum_{j=0}^\infty \frac{u^j}{\Gamma(\rho j+\beta)},
	\end{equation}
	is the two--parametric Mittag--Leffler (ML) function, $\rho,\beta>0$, see also \cite{gorenflo},
	\item for long times
	\begin{equation}\label{eqIIb4}
	R(t;\tau)=-\frac{\tau}{t^{1+\kappa}}\tilde{E}_{-\kappa,-\kappa}\left(-\frac{\tau}{t^\kappa}\right),
	\end{equation}
	where
	\begin{equation}\label{eqIIb5}
	\tilde{E}_{-\rho,-\beta}(u)=\sum_{j=0}^\infty \frac{u^j}{\Gamma(-\rho j-\beta)},
	\end{equation}
	is a generalization of the ML function to negative parameters, $\rho,\beta>0$.
	The distribution $R(t;\tau)$ has a heavy tail,
	\begin{equation}\label{eqIIb6} 
	R(t\rightarrow\infty;\tau)\approx\frac{-\tau}{\Gamma(-\kappa)t^{1+\kappa}},
	\end{equation} 
	the mean value of delay time is infinite. 
	\end{itemize}
We mention that putting $\kappa=1$ in Eq. (\ref{eqIIb1}) we get $R(t;\tau)={\rm e}^{-t/\tau}/\tau$. Then, the average delay time is equal to $\tau$.

In terms of the Laplace transform the Cattaneo--type subdiffusion equation is
\begin{eqnarray}\label{eqIIb7}
\lefteqn{(\tau s^\kappa +1)\left[s\hat{P}(x,s|x_0)-P(x,0|x_0)\right]}\nonumber\\
 &=&Ds^{1-\alpha}\frac{\partial^2 \hat{P}(x,s|x_0)}{\partial x^2}.
\end{eqnarray}
Using the relation 
\begin{equation}\label{eqIIb8}
\mathcal{L}\Bigg[\frac{^C\partial^\alpha f(t)}{\partial t^\alpha}\Bigg](s)=s^\alpha\hat{f}(s)-\sum_{i=0}^{n-1}s^{\alpha-1-i}f^{(i)}(0),
\end{equation}
where 
\begin{equation}\label{eqIIb9}
\frac{^Cd^{\alpha} f(t)}{dt^\alpha}=\frac{1}{\Gamma(n-\alpha)}\int_0^t (t-t')^{n-\alpha-1}f^{(n)}(t')dt',
\end{equation}
$\alpha>0$, is the Caputo fractional derivative (the natural number $n$ is defined as under Eq. (\ref{eqIIa2})), and assuming
\begin{equation}\label{eqIIb10} 
P^{(1)}(x,0|x_0)=0,
\end{equation}
we get
\begin{eqnarray}\label{eqIIb11}
  \lefteqn{\tau\frac{^C\partial^{1+\kappa} P(x,t|x_0)}{\partial t^{1+\kappa}}+\frac{\partial P(x,t|x_0)}{\partial t}}\nonumber\\
  &=&D\frac{^{RL}\partial^{1-\alpha}}{\partial t^{1-\alpha}}\frac{\partial^2 P(x,t|x_0)}{\partial x^2}.
\end{eqnarray}
The above equation can be written in the form
\begin{eqnarray}\label{eqIIb12}
  \lefteqn{\tau\frac{^C\partial^{\alpha+\kappa} P(x,t|x_0)}{\partial t^{1+\kappa}}+\frac{^C\partial^\alpha P(x,t|x_0)}{\partial t^\alpha}}\nonumber\\
  &=&D\frac{\partial^2 P(x,t|x_0)}{\partial x^2}.
\end{eqnarray}

\subsection{Green's functions for an unbounded system\label{secIIc}}

The Green's function is the solution to diffusion equation with the initial conditions
\begin{equation}\label{eqIIc1}
P(x,0|x_0)=\delta(x-x_0), 
\end{equation}
and Eq. (\ref{eqIIb10}).
For the unbounded system, the boundary conditions are
\begin{equation}\label{eqIIc2}
P(\pm\infty,t|x_0)=0.
\end{equation} 
In terms of the Laplace transform the Green's function for Eqs. (\ref{eqIIb11}) and (\ref{eqIIb12}) is 
\begin{equation}\label{eqIIc3}
\hat{P}(x,s|x_0)=\frac{\sqrt{1+\tau s^\kappa}}{2\sqrt{D}s^{1-\alpha/2}}{\rm e}^{-\frac{s^{\alpha/2}|x-x_0|}{\sqrt{D}}\sqrt{1+\tau s^\kappa}}.
\end{equation}
Eq. (\ref{eqIIc3}) takes the form of Eq. (13) in Ref. \cite{gorska2020}.
In the limits of large $s$ and small $s$ (which corresponds to the inverse Laplace transforms for short and long time, respectively) the inverse Laplace transform is determined by applying different procedures. The functions and their Laplace transforms describing the process in the short-time limit will be denoted by the index ST, and the functions for the long-time limit by the index LT.
In the following we use the formula Ref. \cite{tkoszt2004}
\begin{eqnarray}\label{eqIIc4}
  \lefteqn{\mathcal{L}^{-1}\left[s^\nu {\rm e}^{-as^\mu}\right](t)\equiv f_{\nu,\mu}(t;a)}\nonumber\\
&=&\frac{1}{t^{1+\nu}}\sum_{j=0}^\infty \frac{1}{j!\Gamma(-\nu-\mu j)}\left(-\frac{a}{t^\mu}\right)^j,
\end{eqnarray}
$a,\mu>0$, $f_{\nu,\beta}$ is a special case of the Fox H--function,
\begin{eqnarray*}
  \lefteqn{f_{\nu,\mu}(t;a)}\\
  &&=\frac{1}{\mu a^{(1+\nu)/\mu}} H^{1 0}_{1 1}\left(\left.\frac{a^{1/\mu}}{t}\right|
    \begin{array}{cc}
      1 & 1 \\
      (1+\nu)/\mu & 1/\mu
    \end{array}
  \right)
  \;.
\end{eqnarray*}	
In this paper the 50 leading terms of the series in Eq. (\ref{eqIIc4}) are used to calculate the value of the function $f_{\nu,\mu}$.

\subsubsection{Green's function in the limit of short time}

In the limit of large $s$ Eq. (\ref{eqIIc3}) reads
\begin{equation}\label{eqIIc5}
\hat{P}_{ST}(x,s|x_0)=\frac{\sqrt{\tau}}{2\sqrt{D}s^{1-(\alpha+\kappa)/2}}{\rm e}^{-\frac{\sqrt{\tau}s^{(\alpha+\kappa)/2}|x-x_0|}{\sqrt{D}}}.
\end{equation}

Eqs. (\ref{eqIIc4}) and (\ref{eqIIc5}) provide
\begin{eqnarray}\label{eqIIc6}
  \lefteqn{P_{ST}(x,t|x_0)}\nonumber\\
  &=&\frac{\sqrt{\tau}}{2\sqrt{D}}f_{-1+(\alpha+\kappa)/2,(\alpha+\kappa)/2}\left(t;\frac{|x-x_0|\sqrt{\tau}}{\sqrt{D}}\right).\nonumber\\
  &&
\end{eqnarray}

\subsubsection{Green's function in the limit of long time}

When $\tau s^\kappa<1$ a convenient method to determine the inverse Laplace transform of a function $\hat{P}$ is to represent it as a series 
\begin{eqnarray}\label{eqIIc7}
  \hat{P}_{LT}(x,s|x_0)&=&\frac{1}{2\sqrt{D}s^{1-\alpha/2}}{\rm e}^{-\frac{|x-x_0|s^{\alpha/2}}{\sqrt{D}}} \nonumber\\
  &&\times\sum_{i=0}^\infty (\tau s^\kappa)^i a_i\left(\frac{|x-x_0|s^{\alpha/2}}{\sqrt{D}}\right),\nonumber\\
  &&
\end{eqnarray}
and then apply Eq. (\ref{eqIIc4}) term by term.
Applying the series $\sqrt{1+u}=1+u/2-u^2/8+u^3/16-\ldots$, $|u|<1$, and ${\rm e}^{-u}=\sum_{n=0}^\infty (-1)^n u^n/n!$,
for $0\leq i\leq 3$ we get
$a_0(u)=1$, $a_1(u)=(1-u)/2$, $a_2(u)=(-1-u+u^2)/8$, and $a_3(u)=(1+u-u^3/3)/16$.
Keeping the leading terms in the limit $s\rightarrow 0$ we obtain
\begin{eqnarray}\label{eqIIc8}
  \lefteqn{\hat{P}_{LT}(x,s|x_0)}\nonumber\\
  &=&\frac{1}{2\sqrt{D}s^{1-\alpha/2}}{\rm e}^{-\frac{s^{\alpha/2}|x-x_0|}{\sqrt{D}}}\left[1+\frac{\tau}{2}s^\kappa\right.\nonumber\\
    &&\quad\left.-\frac{|x-x_0|\tau}{2\sqrt{D}}s^{\alpha/2+\kappa}-\frac{|x-x_0|\tau^2}{8\sqrt{D}}s^{\alpha/2+2\kappa}\right].\nonumber\\
\end{eqnarray}
From Eqs. (\ref{eqIIc4}) and (\ref{eqIIc8}) we get
\begin{eqnarray}\label{eqIIc9}
  \lefteqn{P_{LT}(x,t|x_0)}\nonumber\\
  &=&\frac{1}{2\sqrt{D}}\left[f_{-1+\alpha/2,\alpha/2}\left(t;\frac{|x-x_0|}{\sqrt{D}}\right)\right.\nonumber\\
&&+\frac{\tau}{2}f_{-1+\kappa+\alpha/2,\alpha/2}\left(t;\frac{|x-x_0|}{\sqrt{D}}\right)\nonumber\\
&&-\frac{|x-x_0|\tau}{2\sqrt{D}}f_{-1+\kappa+\alpha,\alpha/2}\left(t;\frac{|x-x_0|}{\sqrt{D}}\right)\nonumber\\
&&\left.-\frac{|x-x_0|\tau^2}{8\sqrt{D}}f_{-1+2\kappa+\alpha/2,\alpha/2}\left(t;\frac{|x-x_0|}{\sqrt{D}}\right)\right].\nonumber\\
\end{eqnarray}

\section{How the Cattaneo effect changes subdiffusion?\label{secIII}}

The question arises whether the Cattaneo subdiffusion equation brings a new quality in comparison with the ordinary subdiffusion equation. We compare the Green's functions derived for $\tau=0$ and for $\tau\neq 0$, we do the same with the temporal evolutions of mean square particle displacement $\sigma^2$. In this article, the values of parameters and variables are given in arbitrarily chosen units.

\subsection{Influence of the Cattaneo effect on Green's function\label{secIIIa}}

Putting $\tau=0$ in Eq. (\ref{eqIIc3}), then using Eq. (\ref{eqIIc4}) we get the Green's function for the subdiffusion equation, Eq. (\ref{eqIIa1}), without the Cattaneo effect,
\begin{equation}\label{eqIIIa1}
P_{\tau=0}(x,t|x_0)=\frac{1}{2\sqrt{D}}f_{-1+\alpha/2,\alpha/2}\left(t;\frac{|x-x_0|}{\sqrt{D}}\right).
\end{equation}

\begin{figure}[htb]
\centering{%
\includegraphics[width=0.9\linewidth]{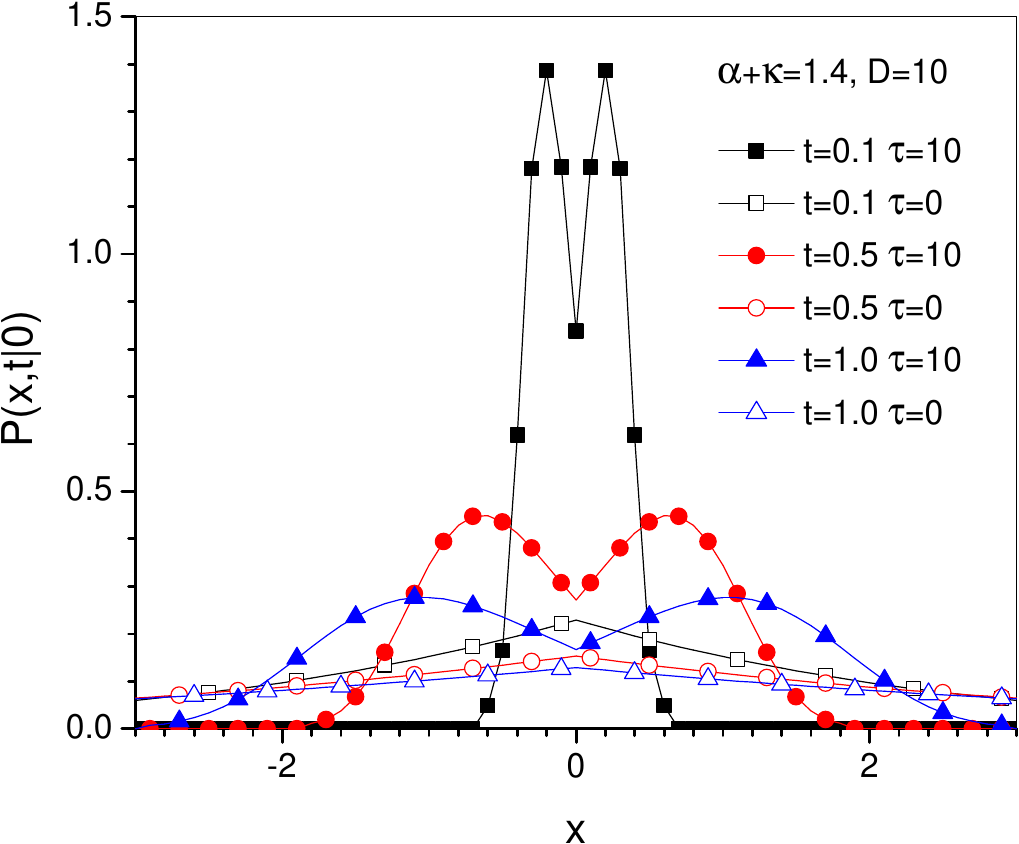}}
\caption{Plots of Green's functions for short times given in the legend. Lines with filled symbols represent Green's functions Eq. (\ref{eqIIc6}) for the subdiffusion equation with the Cattaneo effect, for $\alpha=0.5$, $\kappa=0.9$, $\tau=10$, and $D=10$. Lines with empty symbols are GF Eq. (\ref{eqIIIa1}) plots for the subdiffusion equation without this effect, for $\alpha=0.5$ and $D=10$.}
\label{fig1}
\end{figure}

\begin{figure}[htb]
\centering{%
\includegraphics[width=0.9\linewidth]{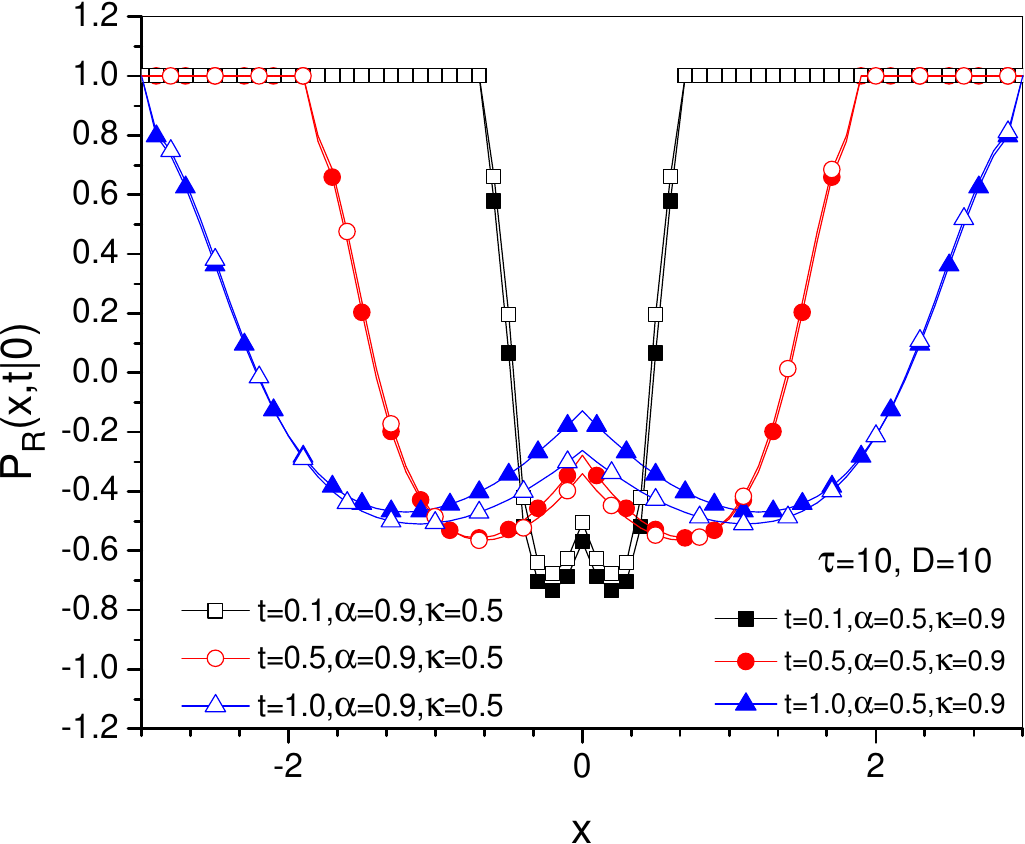}}
\caption{Relative changes of the GF $P_R$ Eq. (\ref{eqIIIa2}) for the functions presented in Fig. \ref{fig1}, for parameters given in the legend.}
\label{fig2}
\end{figure}

\begin{figure}[htb]
\centering{%
\includegraphics[width=0.9\linewidth]{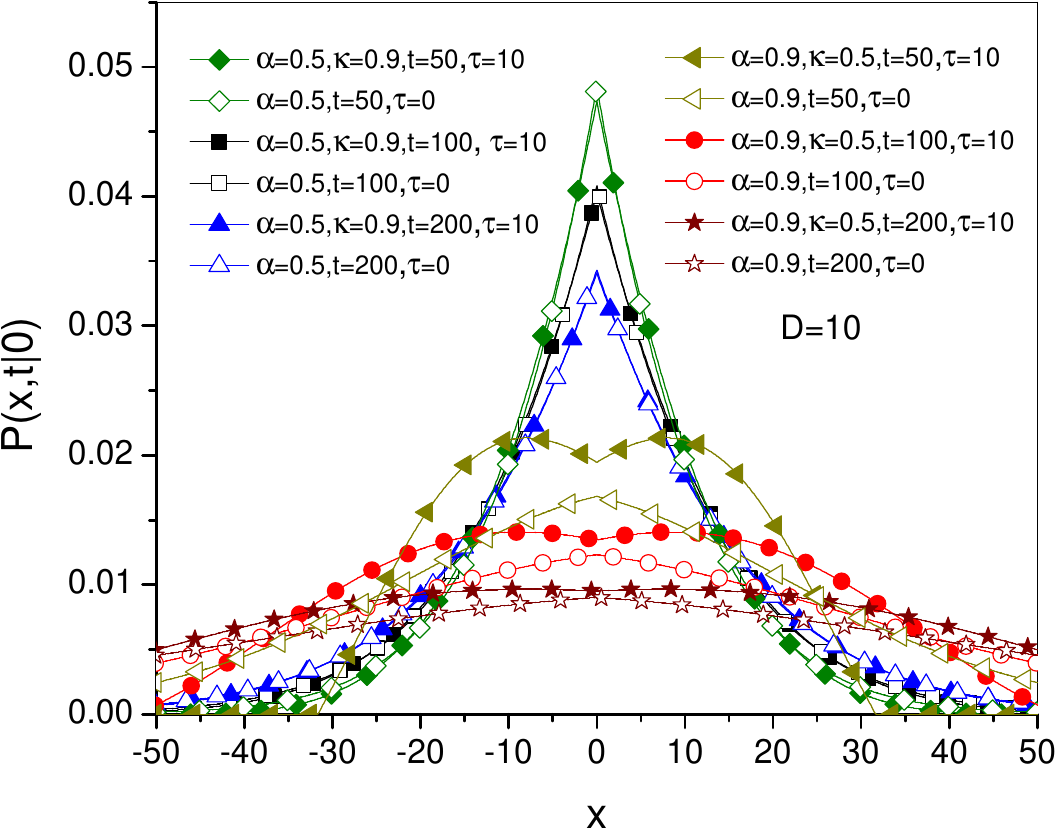}}
\caption{Green's functions for long times Eqs. (\ref{eqIIc9}) and (\ref{eqIIIa1}), the description is analogous to that of Fig. \ref{fig1}.}
\label{fig3}
\end{figure}

\begin{figure}[htb]
\centering{%
\includegraphics[width=0.9\linewidth]{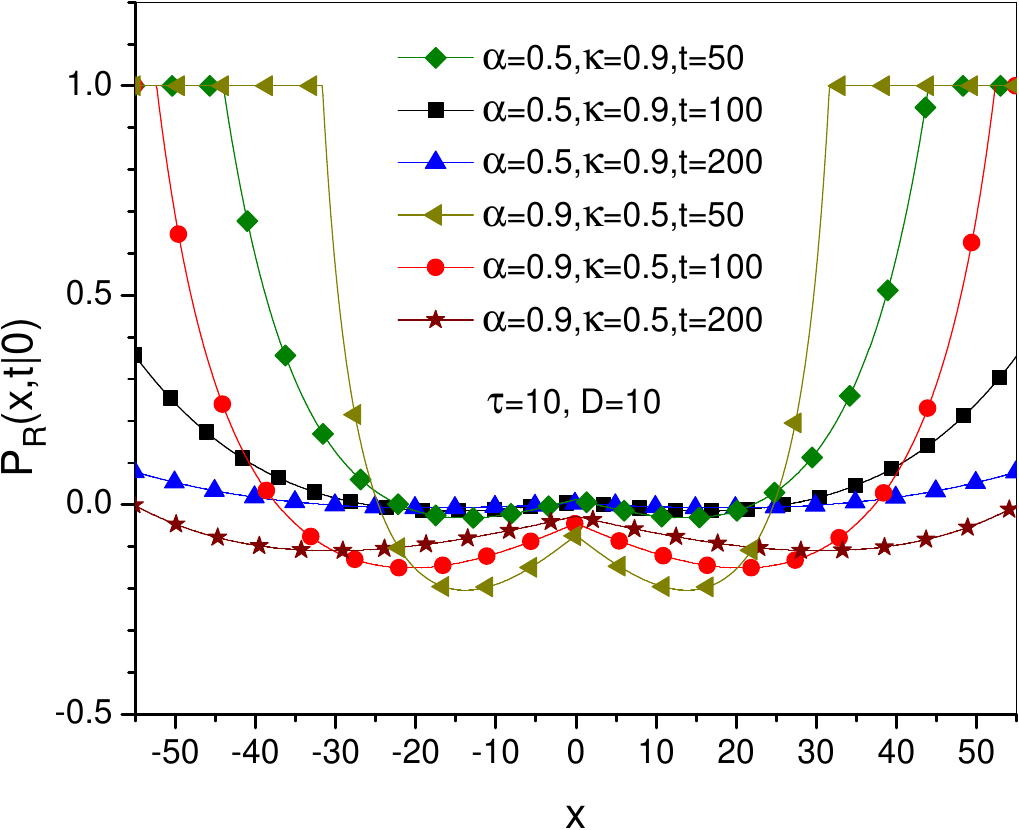}}
\caption{Plots of $P_R$ for long times, the description is analogous to that of Fig. \ref{fig2}.}
\label{fig4}.
\end{figure}

In Figs. \ref{fig1} and \ref{fig3} we compare the plots of functions $P$ and $P_{\tau=0}$ for short and long times, respectively.
Additionally, we consider the relative difference of the Green's functions defined as
\begin{equation}\label{eqIIIa2}
P_R(x,t|x_0)=\frac{P_{\tau=0}(x,t|x_0)-P(x,t|x_0)}{P_{\tau=0}(x,t|x_0)+P(x,t|x_0)},
\end{equation}
the plots of $P_R$ are presented in Figs. \ref{fig2} and \ref{fig4}.

In the short time limit, the GFs do not depend separately on the parameters $\kappa$ and $\alpha$, but on their sum. Two maxima of GFs are observed, which decrease and move away from the particle initial position with time, see Fig. \ref{fig1}. This effect is qualitatively similar to that observed for the GF of the Cattaneo normal diffusion equation Ref. \cite{masweiss}. Fig. \ref{fig2} shows the relative differences between the GFs without and with the Cattaneo effect, the parameters $\alpha$ and $D$ are the same for both functions. We note that the GFs for the Cattaneo effect are significantly different from zero in a narrow interval that widens with time. The relative differences between the GFs depend mainly on the sum $\alpha+\kappa$, and are practically the same for different parameters $\alpha$ which controls $P_{\tau=0}$. Plots of GFs for long times are presented in Figs. \ref{fig3} and \ref{fig4}. In the vicinity of the initial particle position, the differences between GFs without and with the Cattaneo effect disappear with time. 

In Figs. \ref{fig2} and \ref{fig4} there is a region where $P_{R}\approx 1$, which means that $P\approx 0$. The region where the latter function differs noticeably from zero grows with time. This property is qualitatively consistent with the property of the process described by the Cattaneo normal diffusion equation in which the Green's function carrier is finite and grows with time; this property ensures finite velocity of particle propagation.

\subsection{Influence of the Cattaneo effect on temporal evolution of MSD\label{secIIIb}}

As mentioned, a frequently used function that characterizes diffusion is the temporal evolution of the mean square displacement $\sigma^2$ of diffusing particle,
\begin{equation}\label{eqIIIb1}
\sigma^2(t)=\int_{-\infty}^\infty (x-\left\langle x\right\rangle)^2 P(x,t|x_0)dx,
\end{equation}
where $\left\langle x\right\rangle=\int_{-\infty}^\infty xP(x,t|x_0)dx$.
In subdiffusion without the Cattaneo effect there is 
\begin{equation}\label{eqIIIb2}
\sigma^2_{\tau=0}(t)=\frac{2D}{\Gamma(1+\alpha)}t^\alpha.
\end{equation}
From Eq. (\ref{eqIIc3}) and the Laplace transform of Eq. (\ref{eqIIIb1}) we get
\begin{equation}\label{eqIIIb3}
\hat{\sigma^2}(s)=\frac{2D}{s^{1+\alpha}(1+\tau s^\kappa)}.
\end{equation}

\begin{figure}[htb]
\centering{%
\includegraphics[width=0.9\linewidth]{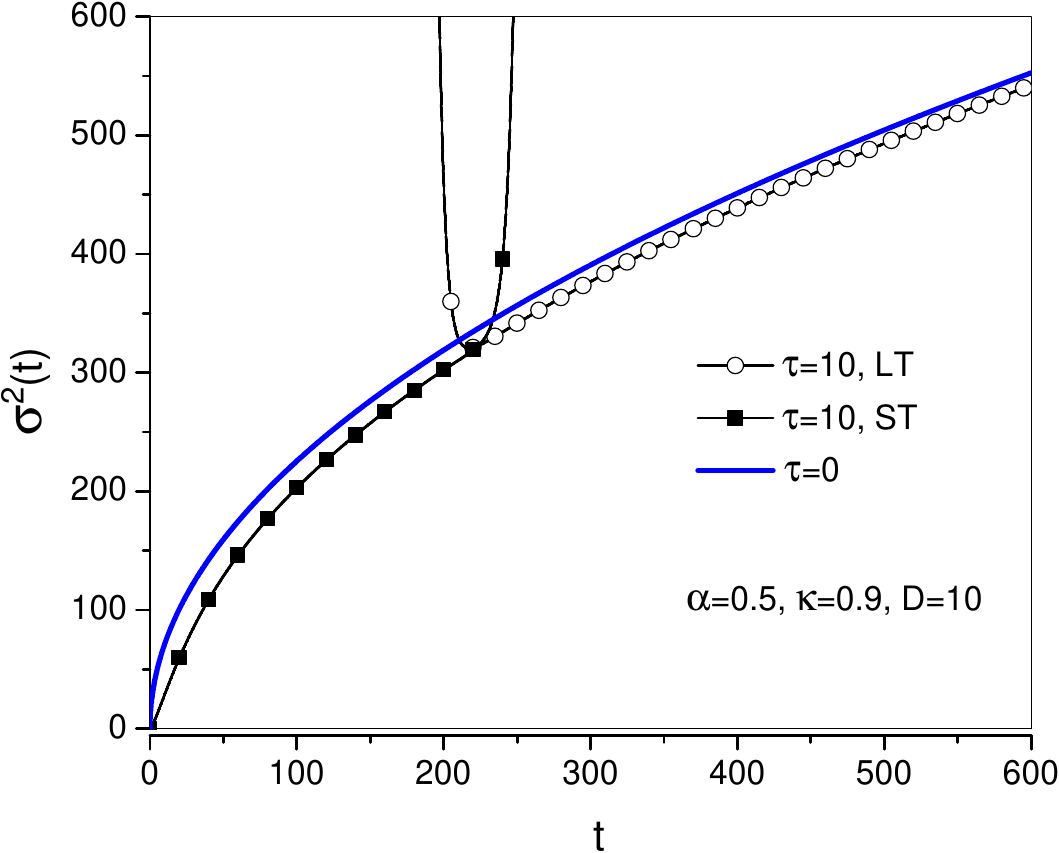}}
\caption{Time evolution of MSD for $\alpha=0.5$ and $\kappa=0.9$. The line with filled symbols is the plot of $\sigma^2_{ST}(t)$ for short times Eq. (\ref{eqIIIb4}), the line with open symbols represents $\sigma^2_{LT}(t)$ for long times Eq. (\ref{eqIIIb5}). The calculations have been made taking into account the 50 leading terms in the series defining the functions. The boundary between short and long times is at the common point of both lines at $t\approx 220$. The line without symbols represents the time evolution of the MSD for the equation without the Cattaneo effect Eq. (\ref{eqIIIb2}).}
\label{fig5}
\end{figure}

\begin{figure}[htb]
\centering{%
\includegraphics[width=0.9\linewidth]{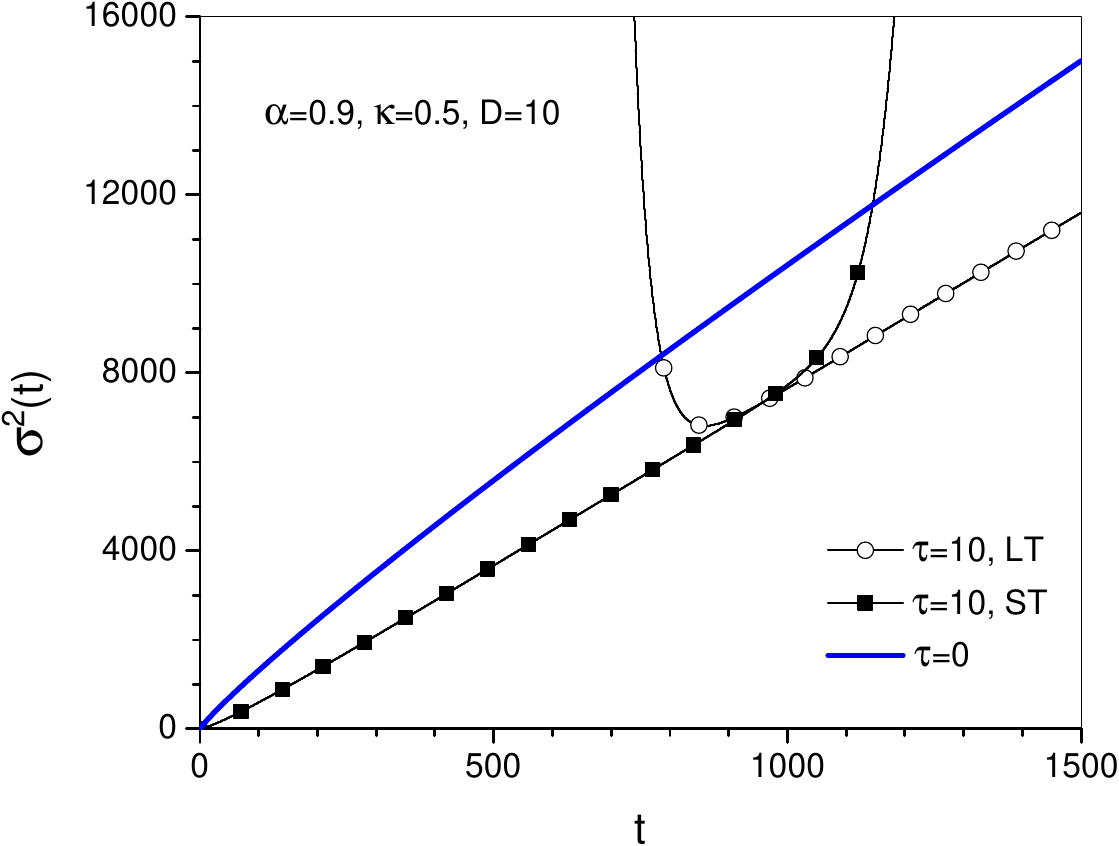}}
\caption{Plots of the time evolution of MSD for $\alpha=0.9$ and $\kappa=0.5$, the description is as for Fig. \ref{fig5}.}
\label{fig6}
\end{figure}

\begin{figure}[htb]
\centering{%
\includegraphics[width=0.9\linewidth]{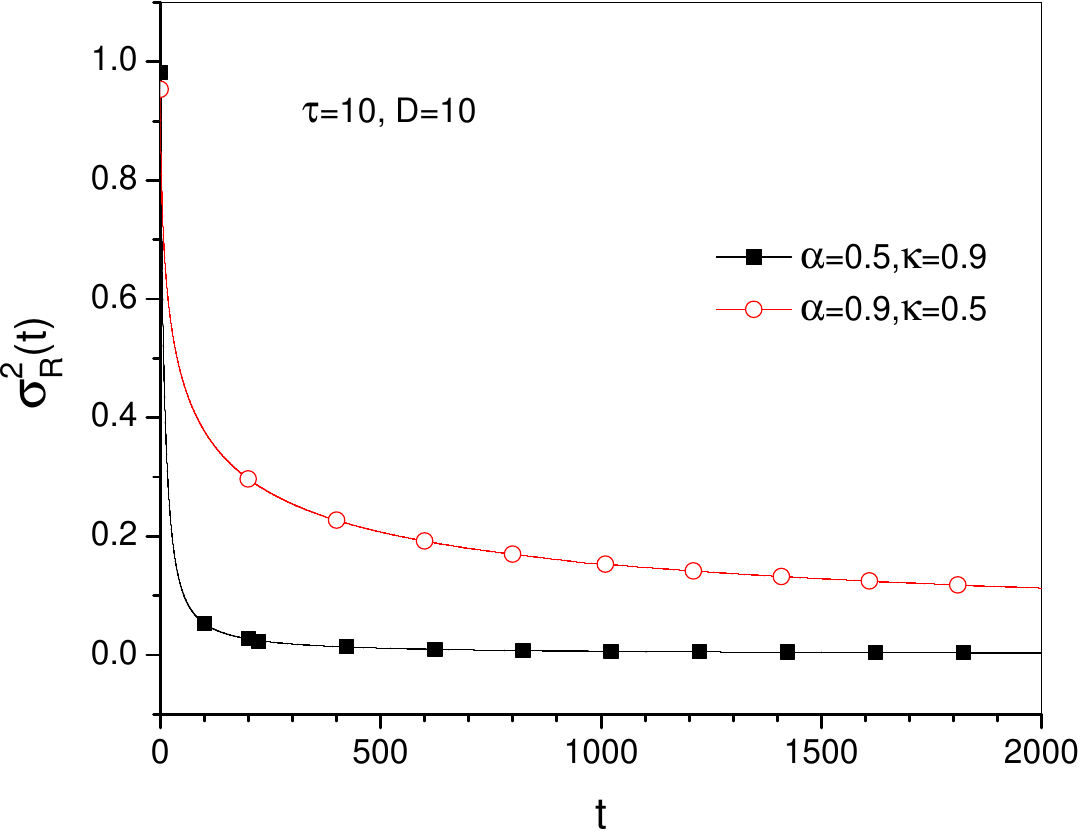}}
\caption{Plots of the functions $\sigma^2_R(t)$ Eq. (\ref{eqIIIb7}) for the parameters given in the legend.}
\label{fig7}
\end{figure}

The inverse Laplace transform method described in Appendix provides for short time
\begin{equation}\label{eqIIIb4}
\sigma^2_{ST}(t)=\frac{2Dt^{\alpha+\kappa}}{\tau}\sum_{i=0}^\infty \frac{(-1)^i}{\Gamma(1+\alpha+(i+1)\kappa)}\left(\frac{t^\kappa}{\tau}\right)^i,
\end{equation}
and for long time
\begin{equation}\label{eqIIIb5}
\sigma^2_{LT}(t)=2Dt^\alpha\sum_{i=0}^\infty \frac{(-\tau)^i}{\Gamma(1+\alpha-i\kappa)t^{i\kappa}}.
\end{equation}

Figures \ref{fig5} and \ref{fig6} show that the time evolution of the MSD for subdiffusion with the Cattaneo effect is smaller than without this effect. Therefore, the Cattaneo effect does not generate superdiffusion despite that for a short time $\sigma^2_{ST}(t\rightarrow 0)=2Dt^{\alpha+\kappa}/[\tau\Gamma(1+\alpha+\kappa)]$. The above equation and Eq. (\ref{eqIIIb2}) provide
\begin{equation}\label{eqIIIb6}
t\ll(\tau\Gamma(1+\alpha+\kappa)/\Gamma(1+\alpha))^{1/\kappa}\implies \sigma^2_{ST}(t)<\sigma^2_{\tau=0}(t).
\end{equation}
Thus, for short times the subdiffusion process with the Cattaneo effect is slower than the subdiffusion without this effect. This follows from the obvious property of the power function that for appropriately small arguments the function values are larger for smaller exponents. As can be seen in Figs. \ref{fig2} and \ref{fig4}, the Cattaneo effect disappears faster when $\alpha<\kappa$ than for the reverse relation.

The relative difference in the time evolution of MSD for processes without and with the Cattaneo effect is defined as
\begin{equation}\label{eqIIIb7}
\sigma^2_R(t)=\frac{\sigma^2_{\tau=0}(t)-\sigma^2(t)}{\sigma^2_{\tau=0}(t)+\sigma^2(t)}.
\end{equation}
Example plots of $\sigma^2_R$ are shown in Fig. \ref{fig7}. In the long time limit there is 
\begin{equation}\label{eqIIIb8}
\sigma^2_R(t\rightarrow\infty)=B_{\alpha,\kappa}\frac{\tau}{t^\kappa},
\end{equation}
where $B_{\alpha,\kappa}=\Gamma(1+\alpha)/[2\Gamma(1+\alpha-\kappa)]$.
The Cattaneo effect disappears in the limit of long time as $1/t^\kappa$.

\section{Cattaneo effect for subdiffusion in a system with partially absorbing wall\label{secIV}} 

The normal and anomalous diffusion equations are derived from a random walk model of a particle. When the particle diffuses through a thin partially permeable or partially absorbing wall, the boundary conditions at the wall can also be derived from this model, taking into account the selective properties of the wall, see the discussion in Refs. \cite{tkoszt2017,tkoszt2015,tkoszt2019}, see also Refs. \cite{greb2019,greb2020,greb2010}. When subdiffusion with Cattaneo effect is considered, the delay in activation of the flux should be also involved in the boundary condition at the wall. We consider subdiffusion with the Cattaneo effect occurring in a system with a partially absorbing wall (PAW); the flux is explicitly involved in the boundary condition at this wall. We determine the Green's functions for this process and the temporal evolution of probability density $W$ of particle absorption by the wall. 

We consider subdiffusion in the interval $x\in(-\infty,0)$ with a partially absorbing wall located at $x=0$.
A frequently used boundary condition at the PAW assumes that the flux is proportional to the concentration of diffusing molecules at the wall. For the Green's function this condition is
\begin{equation}\label{eqIV1}
J(x,t|x_0)=\beta P_W(x,t|x_0),
\end{equation}
where $\beta$ is a parameter controlling the absorption properties of the wall, the index $W$ denotes the Green's function for the system with PAW. When $\beta=0$ the wall is impermeable to diffusing molecules, when $\beta=\infty$ the wall is fully absorbing.
Since the Cattaneo effect delays the activation of the flux, the boundary condition for the Cattaneo-type process, given in terms of the Laplace transform, is as follows
\begin{equation}\label{eqIV2}
-\hat{R}(s;\tau)Ds^{1-\alpha}\frac{\partial \hat{P}_W(x,s|x_0)}{\partial x}\Bigg|_{x=0}=\beta \hat{P}_W(0,s|x_0).
\end{equation}
We assume that $\beta\neq 0$. For $R$ defined by Eq. (\ref{eqIIb1}), in the time domain the boundary condition is
\begin{eqnarray}\label{eqIV3}
  \lefteqn{\left.-D\frac{^{RL}\partial^{1-\alpha}}{\partial t^{1-\alpha}}\frac{\partial P_W(x,t|x_0)}{\partial x}\right|_{x=0}}\nonumber\\
    &&=\beta P_W(0,t|x_0)+\beta\tau\frac{^{RL}\partial^\kappa P_W(0,t|x_0)}{\partial t^\kappa}.
\end{eqnarray}

In terms of the Laplace transform the solution to Eq. (\ref{eqIIb11}) for the initial conditions Eqs. (\ref{eqIIb10}) and (\ref{eqIIc1}), and the boundary conditions Eq. (\ref{eqIV3}) and $P_W(-\infty,t|x_0)=0$ is 
\begin{eqnarray}\label{eqIV4}
  \hat{P}_W(x,s|x_0)=\hat{P}(x,s|x_0)+(1-\Xi(s))\hat{P}(x,s|-x_0),\nonumber\\
\end{eqnarray}
where $x,x_0<0$, $\hat{P}$ is given by Eq. (\ref{eqIIc3}), and $\Xi$ is determined separately for each case considered below.

To characterize the subdiffusion process in the system with PAW, we use the temporal evolution of the absorption probability of a diffusing particle,
\begin{equation}\label{eqIV5}
W(t|x_0)=1-\int_{-\infty}^0 P_W(x,t|x_0)dx.
\end{equation}
The above equations provide
\begin{equation}\label{eqIV6}
\hat{W}(s|x_0)=\frac{1}{2s}\Xi(s){\rm e}^{\frac{-s^\alpha |x_0|\sqrt{1+\tau s^\kappa}}{\sqrt{D}}}.
\end{equation}

\subsection{Equation and boundary condition with the Cattaneo effect\label{secIVa}}

When the Cattaneo effect is included in both the equation and the boundary condition, we have
\begin{eqnarray}\label{eqIV7}
\Xi(s)=\frac{2\beta}{\frac{\sqrt{D}s^{1-\alpha/2}}{\sqrt{1+\tau s^\kappa}}+\beta}.
\end{eqnarray}
As before, the calculation of the Green's function in the time domain is done separately for short and long times. 
For short time we get
\begin{eqnarray}\label{eqIV8}
  \lefteqn{P_{W,ST}(x,t|x_0)}\nonumber\\
  &=&\frac{\sqrt{\tau}}{2\sqrt{D}}\left[f_{-1+(\alpha+\kappa)/2,(\alpha+\kappa)/2}\left(t;\frac{|x-x_0|\sqrt{\tau}}{\sqrt{D}}\right)\right.\nonumber\\
    &&\left.+f_{-1+(\alpha+\kappa)/2,(\alpha+\kappa)/2}\left(t;\frac{|x+x_0|\sqrt{\tau}}{\sqrt{D}}\right)\right]\nonumber\\
  &&-\frac{\beta\tau}{D}f_{-2+\alpha+\kappa,(\alpha+\kappa)/2}\left(t;\frac{|x+x_0|\sqrt{\tau}}{\sqrt{D}}\right),
\end{eqnarray}
\begin{eqnarray}\label{eqIV9}
	W_{ST}(t|x_0)=\frac{\beta\sqrt{\tau}}{\sqrt{D}}f_{-2+(\alpha+\kappa)/2,(\alpha+\kappa)/2}\left(t;\frac{|x_0|\sqrt{\tau}}{\sqrt{D}}\right),
\end{eqnarray}
and in the long time limit there are
\begin{eqnarray}\label{eqIV10}
  \lefteqn{P_{W,LT}(x,t|x_0)}\nonumber\\
  &=&P_{LT}(x,t|x_0)-P_{LT}(x,t|-x_0)\nonumber\\
  &&+\frac{1}{\beta}\left[f_{0,\alpha/2}\left(t;\frac{|x+x_0|}{\sqrt{D}}\right)\right.\nonumber\\
   &&\left. - \frac{\tau}{2\beta}\Big[f_{\kappa,\alpha/2}\left(t;\frac{|x+x_0|}{\sqrt{D}}\right)\right],
\end{eqnarray}	
\begin{eqnarray}\label{eqIV11}
  \lefteqn{W_{LT}(t|x_0)}\nonumber\\
    &=&f_{-1,\alpha/2}\left(t;\frac{|x_0|}{\sqrt{D}}\right)-\frac{\sqrt{D}}{\beta}f_{-\alpha/2,\alpha/2}\left(t;\frac{|x_0|}{\sqrt{D}}\right)\nonumber\\
  &&-\frac{\tau|x_0|}{\sqrt{D}}f_{-1+\alpha/2+\kappa,\alpha/2}\left(t;\frac{|x_0|\sqrt{\tau}}{\sqrt{D}}\right).
\end{eqnarray}

\begin{figure}[htb]
\centering{%
\includegraphics[width=0.9\linewidth]{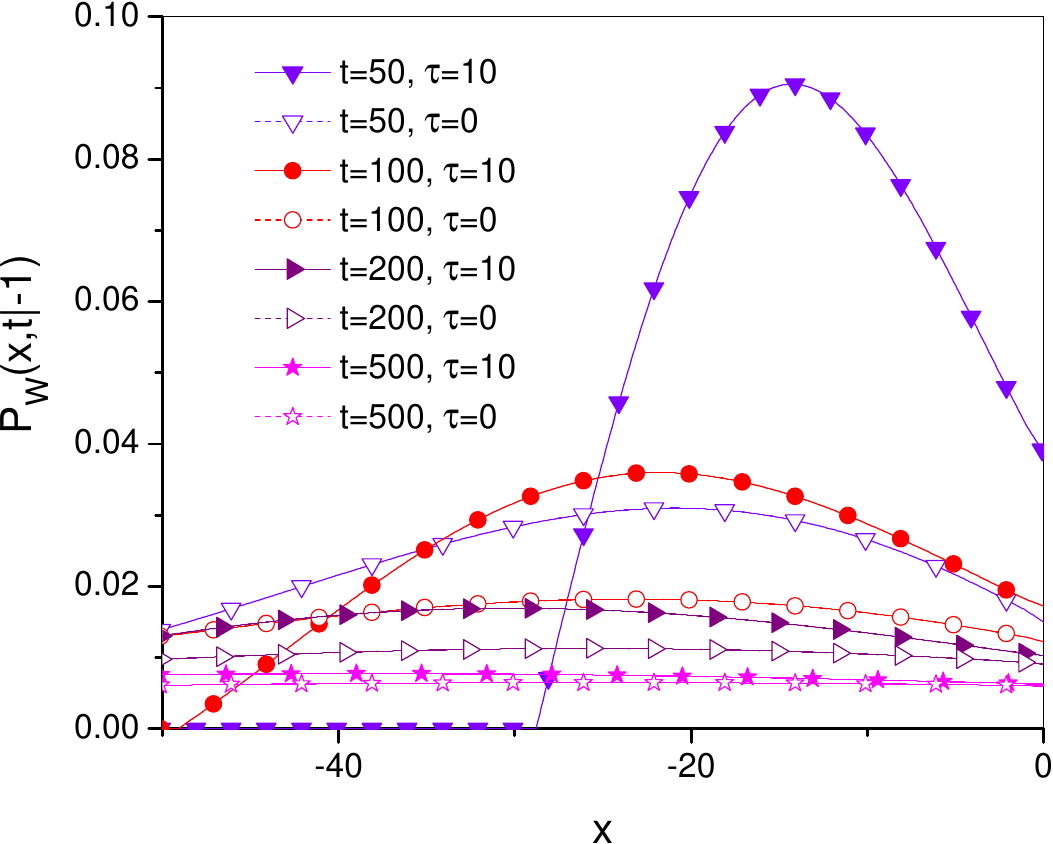}}
\caption{Green's functions for subdiffusion in a system with PAW located at $x=0$, with the Cattaneo effect Eq. (\ref{eqIV10}) (lines with full symbols) and without this effect Eq. (\ref{eqIV15}) (lines with open symbols) for times given in the legend, here $\alpha=0.9$, $\kappa=0.5$, $\beta=0.1$, $x_0=-1$, and $D=10$.}
\label{fig8}
\end{figure}

\begin{figure}[htb]
\centering{%
\includegraphics[width=0.9\linewidth]{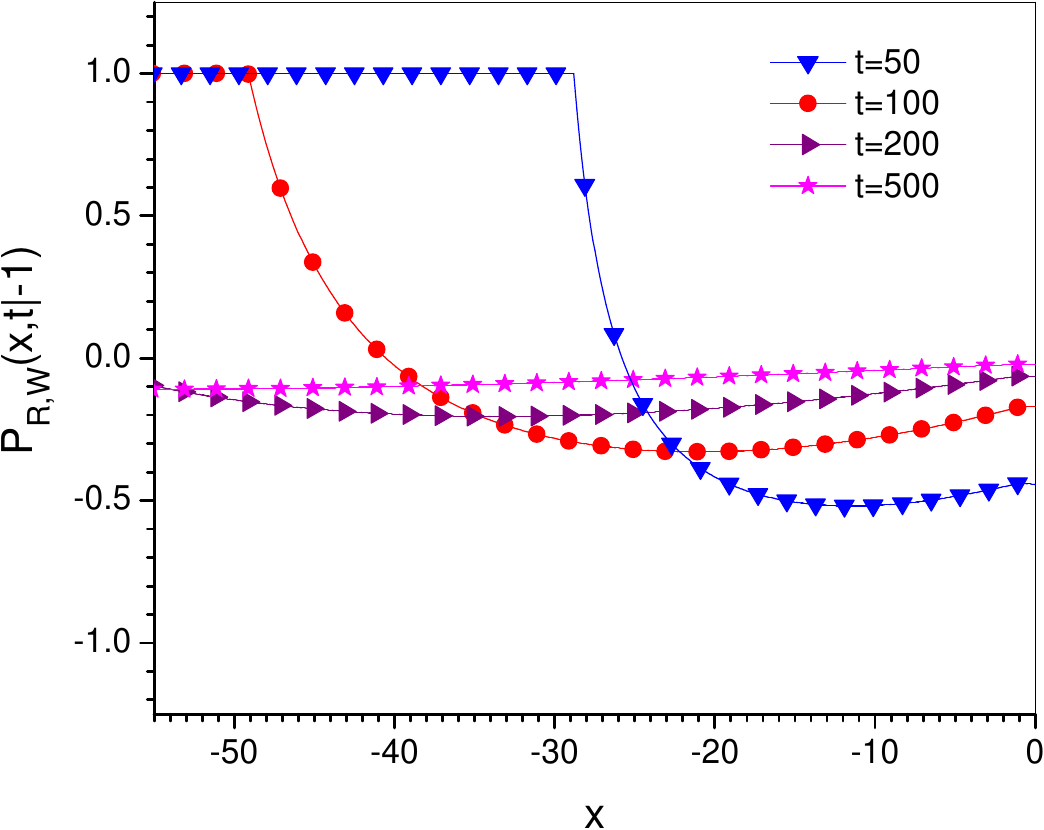}}
\caption{Time evolution of the relative change of the Green's function Eq. (\ref{eqIIIa2}) for the functions presented in Fig. \ref{fig8}.}
\label{fig9}
\end{figure}

\subsection{Equation and boundary condition without the Cattaneo effect\label{secIVb}}

When the equation and boundary condition are without the Cattaneo effect we get
\begin{eqnarray}\label{eqIV12}
\Xi_{\tau=0}(s)=\frac{2\beta}{\sqrt{D}s^{1-\alpha/2}+\beta}.
\end{eqnarray}
For short time the approximations of GF and $W$ read
\begin{eqnarray}\label{eqIV13}
  \lefteqn{P_{W,\tau=0,ST}(x,t|x_0)}\nonumber\\
  &=&\frac{1}{2\sqrt{D}}\left[f_{-1+\alpha/2,\alpha/2}\left(t;\frac{|x-x_0|}{\sqrt{D}}\right)\right.\nonumber\\
    &&\left.+f_{-1+\alpha/2,\alpha/2}\left(t;\frac{|x+x_0|}{\sqrt{D}}\right)\right]\nonumber\\
  &&-\frac{\beta}{D}f_{-2+\alpha,\alpha/2}\left(t;\frac{|x+x_0|}{\sqrt{D}}\right),
\end{eqnarray}
\begin{eqnarray}\label{eqIV14}
	W_{\tau=0,ST}(t|x_0)=\frac{\beta}{\sqrt{D}}f_{-2+\alpha/2,\alpha/2}\left(t;\frac{|x_0|}{\sqrt{D}}\right).
\end{eqnarray}
For long time the functions are approximated as
\begin{eqnarray}\label{eqIV15}
  \lefteqn{P_{W,\tau=0,LT}(x,t|x_0)}\nonumber\\
  &=&P_{LT}(x,t|x_0)-P_{LT}(x,t|-x_0)\nonumber\\
  &&+\frac{1}{\beta}\Big[f_{0,\alpha/2}\left(t;\frac{|x+x_0|}{\sqrt{D}}\right)\nonumber\\
  &&-\frac{\sqrt{D}}{\beta}f_{1-\alpha/2,\alpha/2}\left(t;\frac{|x+x_0|}{\sqrt{D}}\right)\Big],
\end{eqnarray}
\begin{eqnarray}\label{eqIV16}
  \lefteqn{W_{\tau=0,LT}(t|x_0)}\nonumber\\
  &=&f_{-1,\alpha/2}\left(t;\frac{|x_0|}{\sqrt{D}}\right)-\frac{\sqrt{D}}{\beta}f_{-\alpha/2,\alpha/2}\left(t;\frac{|x_0|}{\sqrt{D}}\right)\nonumber\\
  &&-\frac{D}{\beta}f_{-1+\alpha/2+\kappa,\alpha/2}\left(t;\frac{|x_0|\sqrt{\tau}}{\sqrt{D}}\right).
	\end{eqnarray}

Fig. \ref{fig8} shows plots of Green's functions Eqs. (\ref{eqIV10}) and (\ref{eqIV15}). The functions in the system with the Cattaneo effect enabled in the equation and boundary condition are compared with the functions determined in the system without this effect. The relative difference of these functions is presented in Fig. \ref{fig9}. As previously, it can be seen that there is a region where $P_{W,R}=1$, which means that $P_W(x,t|x_0)=0$ for $x$ far away from $x_0$. This fact indicates that the velocity of particle propagation is limited.

\subsection{Can the Cattaneo effect be neglected in the boundary condition?\label{secIVc}}

The question arises whether including the Cattaneo effect in the boundary condition significantly affects the solutions to CTSE. Let us consider subdiffusion in a system with PAW, described by a Cattaneo-type equation, but where this effect is not included in the boundary condition at the wall. In this case the boundary condition reads
\begin{eqnarray}\label{eqIV17}
  \lefteqn{-D\frac{^{RL}\partial^{1-\alpha}}{\partial t^{1-\alpha}}\frac{\partial P_{W,J:\tau=0}(x,t|x_0)}{\partial x}\Bigg|_{x=0}}\nonumber\\
  &=&\beta P_{W,J:\tau=0}(0,t|x_0),
\end{eqnarray}
the functions are marked with the index $J:\tau=0$. We get
\begin{eqnarray}\label{eqIV18}
\Xi_{J:\tau=0}(s)=\frac{2\beta}{\sqrt{D}s^{1-\alpha/2}\sqrt{1+\tau s^\kappa}+\beta}.
\end{eqnarray}
As an example showing that including the Cattaneo effect in the boundary condition significantly changes the subdiffusion process, we consider the function $W$ in the short-time approximation,
\begin{eqnarray}\label{eqIV19}
  \lefteqn{W_{J:\tau=0,ST}(t|x_0)}\nonumber\\
  &=&\frac{\beta}{\sqrt{D\tau}}f_{-2+(\alpha-\kappa)/2,(\alpha+\kappa)/2}\left(t;\frac{|x_0|\sqrt{\tau}}{\sqrt{D}}\right).\nonumber\\
  &&
\end{eqnarray}

\begin{figure}[htb]
\centering{%
\includegraphics[width=0.9\linewidth]{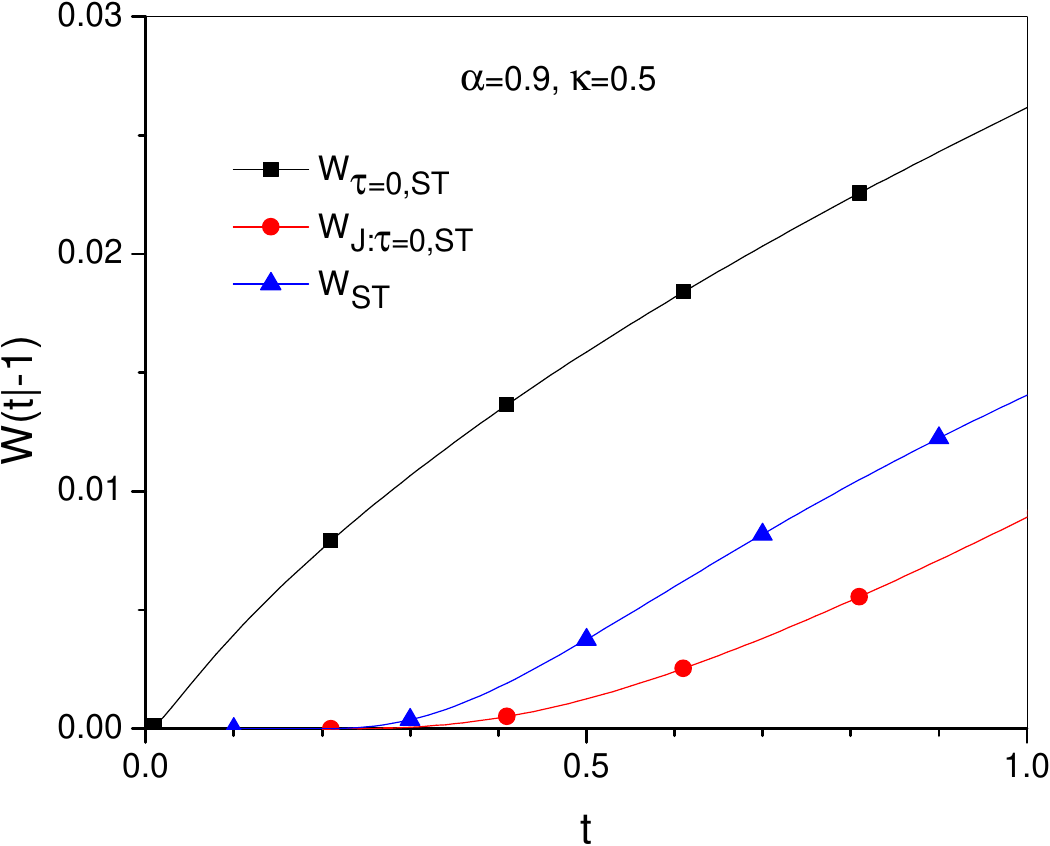}}
\caption{Plots of the functions $W$ for the cases: without the Cattaneo effect for the equation and boundary condition Eq. (\ref{eqIV14}) (marked with squares), with the effect enabled only in the equation Eq. (\ref{eqIV19}) (circles), and with the effect enabled in both the equation and the boundary condition Eq. (\ref{eqIV9}) (triangles), for parameters $\alpha=0.9$, $\kappa=0.5$, $\beta=0.1$, $\tau=1$, $x_0=-1$, and $D=10$.}
\label{fig10}
\end{figure}

\begin{figure}[htb]
\centering{%
\includegraphics[width=0.9\linewidth]{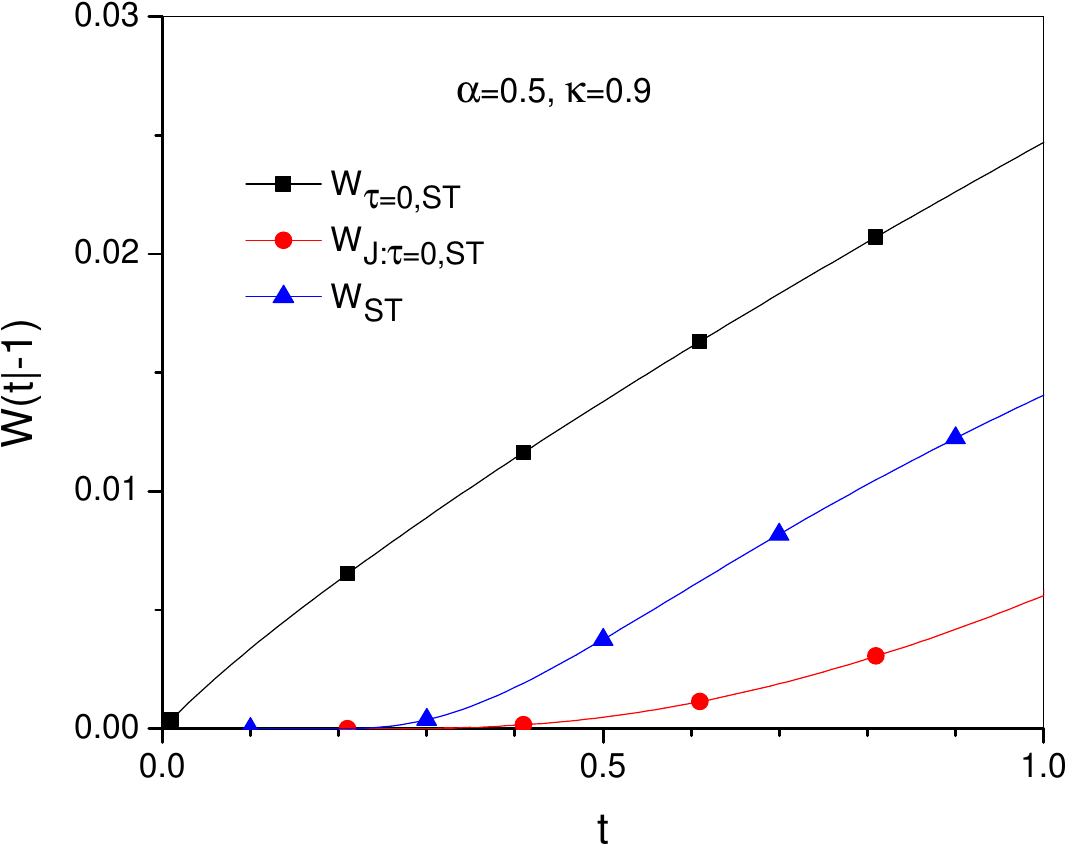}}
\caption{Plots of $W$ for $\alpha=0.5$ and $\kappa=0.9$, description and values of the other parameters are the same as for Fig. \ref{fig10}.}
\label{fig11}
\end{figure}

The plots of $W_{ST}$ are presented in Figs. \ref{fig10} and \ref{fig11}. They show that including the Cattaneo effect in the boundary condition significantly changes the solutions to the Cattaneo-type subdiffusion equation. In our opinion, including the Cattaneo effect only in the equation is unphysical and makes the model inconsistent.

\section{Final remarks\label{secV}}

The Cattaneo-type subdiffusion equation describes subdiffusion with involved Cattaneo effect which consists in delaying the activation of the ordinary subdiffusion flux by a random delay time. 
Anomalous diffusion processes have been described by various differential-integral equations with memory kernel (MK). The examples are equations generated by distributed order MK \cite{sandev2015,sandev2015a,sandev2017,sandev2018,sandev2019}, power--law MK \cite{sandev2015}, and truncated power law MK \cite{sandev2017}, see also \cite{metzler1998,sandev2015a,sandev2018,sandev2019,chechkin2021}. The general form of CTSE, Eq. (\ref{eqIIa11}), is a special case of an equation with MK which is controlled by the probability density of delay time $R$. Then, MK is limited by the conditions 1-3 in Sec. \ref{secIIa} imposed on the function $R$.

The interpretation of the Cattaneo effect is as follows. After making a jump, the particle ability to make the next jump is temporarily delayed. The reasons for this depend on the system in which the particle is randomly moving. One of them is an inertia of the particle. This also applies to a certain delay in making a ``random'' decision to perform an action. The delay is caused by the overload of chaotic information, which can occur in economic processes or in the movement of people in dense crowds.
Subdiffusion typically occurs in highly complex environments, such as the viscoelastic network of chromatin \cite{lee2021}, an agarose gel \cite{kdm}, or a bacterial biofilm \cite{km,kmwa}. In the initial stage, the particle can be immobilized e.g. by momentary deformation of fibers of the matrix. The medium exerts mechanical resistance, it behaves like a stretched spring, trapping the molecule. After the random delay time, the elastic constraints of the matrix give way, the pores expand allowing the particle to jump. 
An example of a process in which the Cattaneo effect may be important is subdiffusion (or normal diffusion when $\alpha=1$) of antibiotics in a bacterial biofilm. The defense mechanisms of bacteria against the action of antibiotics can change the random walk of antibiotic molecules \cite{aot,mot}. Bacterial defense mechanisms involve, among others, hindering the movement of antibiotic molecules within the biofilm. This process can be divided into stages, see Refs. \cite{km,kmwa} and the references cited therein. It is possible that initially the molecules are immobilized, and once the biofilm is weakened, antibiotic subdiffusion within the biofilm begins.

The function $\sigma^2(t)$ is the time-evolving variance of the Green's function, which describes the rate of spread of the diffusing substance. In Sec. \ref{secIIIb} it has been shown that the process described by the Cattaneo-type subdiffusion equation leads to a time evolution of the MSD that is slower than that for the process without the Cattaneo effect, assuming that the parameters $\alpha$ and $D$ are the same for both processes. We conclude that the CTSE equation describes the subdiffusion process in the entire time domain, even though the relation $\sigma^2(t\rightarrow 0)\sim t^\nu$ with $\nu>1$ suggests superdiffusion. 
This leads to the conclusion that $\sigma^2(t)$ does not uniquely define the type of diffusion in the limit of short time, see also Refs. \cite{meroz2011,dybiec1}.

We have argued that when the diffusion process is described by the Cattaneo-type subdiffusion equation, the appearance of a flux in a ``standard'' boundary condition necessitates the inclusion of the Cattaneo effect in that boundary condition as well.
We considered subdiffusion in a system with a partially absorbing wall, where the standard boundary condition at the wall assumes that the flux is proportional to the concentration of the diffusing particles or, when considering single-particle diffusion, to the value of Green's function at the wall. Including the Cattaneo effect in the boundary condition introduces a new term in the condition, controlled by the parameters of the flux activation delay distribution, see Eq. (\ref{eqIV3}).

The question arises how to experimentally identify and measure the Cattaneo effect. It is convenient to study the concentration distribution of a diffusing substance in a system consisting of two vessels separated by a thin partially permeable membrane. The concentration profile is relatively easy to measure experimentally by means of an interferometric method \cite{kdm,kmwa,kwl2017}. 
Assuming the system is homogeneous in planes perpendicular to the $x$ axis, the problem is effectively one-dimensional. Let a membrane separate vessels with known initial concentrations of diffusing substance.
By comparing the time evolution of the amount of substance in one of the vessels with the theoretical function obtained from the solutions of the CTSE, the Cattaneo effect could be identified along with the estimation of parameters of the function $R$.
We mention that the method has been used in the experimental determination of membrane boundary conditions \cite{kwl2017}, as well as in the identification and determination of subdiffusion parameters in gels \cite{kdm} and bacterial biofilms \cite{kmwa}. 
In Ref. \cite{kwl2017} there was found the boundary condition at a thin partially permeable membrane separating two vessels containing aqueous solutions of ethanol. Normal diffusion occurs in the system, but the boundary condition included an additional term with the Riemann-Liouville time fractional derivative of the 1/2 order. This term suggests that the permeation of ethanol through the membrane is a process with a long memory component, which could be generated by the Cattaneo effect. However, the problem of experimental identification of the Cattaneo effect in a membrane system requires further research.

The reason for deriving the Cattaneo equation for normal diffusion was that the equation described a process with a finite maximum propagation velocity of diffusing molecules. As far as we know, there is no rigorous proof that the maximum propagation velocity of a particle for the Cattaneo-type subdiffusion equation is finite. However, numerical solutions of CTSE suggest that a particle propagation velocity is limited. This is visible in Figs. \ref{fig2} and \ref{fig4}, where plots of the function $P_R$ are presented. Since $P_R\rightarrow 1$ when $P\rightarrow 0$, the plots suggest that the region over which $P$ is different from zero is finite and increases with time.

The parameters $\alpha$ and $\kappa$, controlling the orders of time fractional derivatives in the CTSE, are assumed to be independent of each other. This is a more universal model than models assuming that the fractional derivatives in the Cattaneo equation are controlled by the parameter $\alpha$ only. An example is the model considered in Refs. \cite{pietrzak}, where the Cattaneo subdiffusion equation contains fractional time derivatives of orders $\alpha$ and $2\alpha$.

\section*{Appendix: Calculating the inverse Laplace transform}

We describe the method used in this paper to determine the inverse Laplace transforms of certain functions. We do not directly determine the inverse transform $\mathcal{L}^{-1}[\hat{g}(s)](t)$, but the transform $\mathcal{L}^{-1}[{\rm e}^{-as^\mu}\hat{g}(s)](t)$, $a,\mu>0$, for which we apply Eq. (\ref{eqIIc4}) and then calculate the limit $a\rightarrow 0^+$; the result is independent of the parameter $\mu$. For example, suppose $\hat{g}(s)$ has two different power series expansions for large and small values of the positive real parameter $s$,
\begin{equation}\label{d1}
\hat{g}(s)=\frac{1}{s^{c_1}}\sum_{n=0}^\infty\frac{a_n}{s^{c_2 n}},\;s>s_1,
\end{equation}
\begin{equation}\label{d2}
\hat{g}(s)=s^{d_1}\sum_{n=0}^\infty b_n s^{d_2 n},\;s<s_2,
\end{equation}
where $c_2,d_2>0$ and $s_1$, $s_2$ are positive numbers.
Let us consider the following series
\begin{equation}\label{d3}
{\rm e}^{-as^\mu}\hat{g}(s)=\frac{{\rm e}^{-as^\mu}}{s^{c_1}}\sum_{n=0}^\infty\frac{a_n}{s^{c_2 n}},\;s>s_1,
\end{equation}
\begin{equation}\label{d4}
{\rm e}^{-as^\mu}\hat{g}(s)={\rm e}^{-as^\mu} s^{d_1}\sum_{n=0}^\infty b_n s^{d_2 n},\;s<s_2,
\end{equation}
$a,\mu>0$.
Eqs. (\ref{eqIIc4}) and (\ref{d3}) give
\begin{equation}\label{d5}
\mathcal{L}^{-1}[{\rm e}^{-as^\mu}\hat{g}(s)](t)=\sum_{n=0}^\infty a_n f_{-c_1-c_2n,\mu}(a;t),\;t<t_1,
\end{equation}
and Eqs. (\ref{eqIIc4}) and (\ref{d4}) provide
\begin{equation}\label{d6}
\mathcal{L}^{-1}[{\rm e}^{-as^\mu}\hat{g}(s)](t)=\sum_{n=0}^\infty b_n f_{d_1+d_2n,\mu}(a;t),\;t>t_2.
\end{equation}
In the limit $a\rightarrow 0^+$ we have $f_{\nu,\mu}(t;a)\rightarrow 1/[\Gamma(-\nu)t^{1+\nu}]$; $f_{n,\mu}(t;a)=0$ when $\nu=n$ is a natural number. When $a\rightarrow 0^+$, Eqs. (\ref{d5}) and (\ref{d6}) are, respectively, 
\begin{equation}\label{d7}
g(t)=\sum_{n=0}^\infty a_n \frac{t^{c_1+c_2 n-1}}{\Gamma(c_1+c_2 n)},\;t<t_1,
\end{equation}
\begin{equation}\label{d8}
g(t)=\sum_{n=0}^\infty b_n \frac{1}{\Gamma(-d_1-d_2 n)t^{d_1+d_2 n+1}},\;t>t_2.
\end{equation}
The inequalities $s>s_1$ and $s<s_2$ do not uniquely determine the corresponding parameters $t_1$ and $t_2$.
Determining these parameters requires additional considerations. One such method is to determine the time intervals $I_1=(0,t_1)$ and $I_2=(t_2,\infty)$ in which the series converge and the functions defined in the second of these intervals can be interpreted as an extension of the function defined in the first interval. Precisely determining the time intervals over which the long- and short-time approximations work would require additional considerations, see the discussion in Ref. \cite{tk2023}. 

\bibliographystyle{cas-model2-names}

%\bibliography{cas-refs}

\end{document}